\documentclass[12pt]{article}

\newcommand{\be}{\begin{equation}}
\newcommand{\ee}{\end{equation}}
\newcommand{\bea}{\begin{eqnarray}}
\newcommand{\eea}{\end{eqnarray}}

\begin{document}

\begin{titlepage}
\begin{center}
\vskip .2in
\hfill
\vbox{
    \halign{#\hfil         \cr
           hep-th/0003163 \cr
           SU-ITP-0010 \cr
           March 2000    \cr
           }  
      }   
\vskip 0.5cm
{\large \bf Raiders of the Lost $AdS$}\\
\vskip .2in
{\bf  Jason Kumar}
\footnote{e-mail address:jkumar@leland.stanford.edu}\\
\vskip .25in
{\em
Department of Physics,
Stanford University, Stanford, California 94305 USA \\}
\vskip 1cm
\end{center}
\begin{abstract}We demonstrate that under certain conditions
a theory of conformal quantum mechanics will exhibit the
symmetries of two half-Virasoro algebras.  We further
demonstrate the conditions under which these algebras
combine to form a single Virasoro algebra, and comment
on the connection between this result and the $AdS_2 /
CFT_1$ correspondence.
\vskip 0.5cm
 \end{abstract}
\end{titlepage}
\newpage

\section{Introduction}

\vskip .4in

$\it {\vbox{
    \halign{#\hfil         \cr
           \hskip 1.8in Indy:\hskip .15in{\it I'm going after
                             that [$AdS$].} \cr
           \hskip 1.8in Sallah:\hskip .03in{\it How?} \cr
           \hskip 1.8in Indy:\hskip .15in {\it I don't know.
                            I'm making this up as I go.}\cr
           \hskip 2.1in ``Raiders of the Lost Ark" \cr
           }  
      }   
      }$

\vskip .4in

Recently, there has been great interest in the
$AdS / CFT$ correspondence (\cite{Mald}-\cite{Witten},
and many others).  Although there has been a great deal
of work related to the study of $AdS_3$, $AdS_4$,
$AdS_5$ and $AdS_7$, there has been relatively little
study of the $AdS_2$ case.  Type II supergravity exhibits
solutions of the form $AdS_2 \times S^2 \times T^6$,
$AdS_2 \times S^3 \times T^5$ and other solutions
related to Calabi-Yau compactification.

On one hand, the $AdS_2 /CFT_1$ correspondence has the
potential to be extremely fascinating.  A weakly
curved $AdS_2 \times S^2$ space will be an approximately
flat 4 dimensional space.  As a result, the
$AdS_2 /CFT_1$ correspondence naively may be able to make
non-trivial statements about 4 dimensional quantum
gravity.  On the other hand, there are potential
obstacles to a true understanding or complete
formulation of the $AdS_2 /CFT_1$ correspondence, such
as the fragmentation of $AdS_2$, etc.

Nevertheless, a clearer picture of the relationship
between $AdS_2$ and $CFT_1$ has recently emerged
\cite{Andy}-\cite{DFG}.  In
particular, it was noted in \cite{Andy} that
any theory of quantum gravity in the background of
$AdS_2$ contains not only an $SL(2,R)$ symmetry, but also
the symmetries of a Virasoro algebra.
This may imply that any boundary theory holographically
dual to $AdS_2$ should also contain the symmetries of
a Virasoro algebra.  In \cite{Kumar}, it was shown
that any classical scale-invariant mechanics of one
variable exhibited not only conformal invariance, but
also the symmetries of a Virasoro algebra.  In a few
instances, this result was extended to a quantum
mechanical statement of the commutation relations between
generators (as opposed to a statment regarding Poisson
bracket relations between generator functions).  However,
a general statement regarding multi-particle systems and
operator algebras was still missing.

In this paper we demonstrate that, under certain conditions,
a theory of conformal quantum mechanics (of an arbitrary
number of variables) exhibits the symmetries of two half-
Virasoro algebras.  If a further condition holds, then these
two algebras can be combined to form a single Virasoro algebra.

\subsection{Classical Generators}

The conformal group in $0+1$ dimensions is
$SL(2,R)$.  The
algebra can be written as
\be
[D,H]=\imath \hbar H \qquad
[D,K]=-\imath \hbar K \qquad
[H,K]=2\imath \hbar D
\ee
If one makes the identification
\be
\label{embedding}
L_{-1}=-{\imath\over  \hbar} H \qquad
L_0 = -{\imath\over  \hbar} D \qquad
L_1 = {\imath\over  \hbar} K,
\ee
then the conformal algebra is identical to the
$SL(2,R)$
algebra formed by the global subalgebra of the Virasoro
algebra (note that there are other ways of embedding
the conformal algebra in the Virasoro algebra).

In \cite{Kumar} it was shown that if the conformal
algebra of a theory of classical conformal mechanics
is embedded in the global subalgebra of the
Virasoro algebra as shown above, then generators of
a full Virasoro algebra can be found.

In the examples discussed in \cite{Kumar}, one can see
that the Virasoro generators can be written in the form
\footnote{We thank Sangmin Lee for making this point.}
\be
L_m = L_0 ^{1+m} L_{-1} ^{-m}.
\ee

One can easily show at the level of Poisson brackets that
if $\{L_0, L_{-1}\}_{PB} = L_{-1}$, then the generators
defined above satisfy the algebra
\bea
\{L_m, L_n\}_{PB} &=& (1+m)(-n)L_0 ^{1+m+n} L_{-1} ^{-m-n-1}
\{L_0, L_{-1}\}_{PB} \nonumber\\
&\,&
+(1+n)(-m) L_0 ^{1+m+n} L_{-1} ^{-m-n-1}
\{L_{-1}, L_0\}_{PB} \nonumber\\
&=& [(1+m)(-n) - (1+n)(-m)]L_{m+n} \nonumber\\
&=& (m-n) L_{m+n}.
\eea

Thus, any system of classical mechanics with scale-invariance
also exhibits the symmetries of a full Virasoro algebra (at the
level of Poisson brackets).  This is a generalization of the
work done in \cite{Kumar}, as it applies to the case
of an arbitrary number of variables.  However, the analysis is
still only classical in nature.  In order to make a statment about
conformal quantum mechanics, one must also understand the issues
related to the normal ordering of operators.

\subsection{Quantum Generators}

Suppose that the generators $L_0$, $L_1$
and $L_{-1}$ satisfy the $SL(2,R)$ algebra
given by
\be
[L_0, L_1] = -L_1 \qquad
[L_0, L_{-1}] = L_{-1} \qquad
[L_1, L_{-1}] = 2L_0.
\ee

If the operator $L_{-1}$ is invertible,
then we may consistently define the
operator $L_{-1} ^{-1}$.  This operator has the commutation
relation $[L_0, L_{-1} ^{-1}] = -L_{-1} ^{-1}$.
We then make the ansatz

\be
L_m = L_0 (L_{-1} ^{-1} L_0 )^m  =
(L_0 L_{-1} ^{-1})^m L_0 \qquad
m\geq 0.
\ee
One can easily see that

\bea
[L_m, L_n]&=& L_0 (L_{-1} ^{-1} L_0 )^m
L_0 (L_{-1} ^{-1} L_0 )^n -
L_0 (L_{-1} ^{-1} L_0 )^n
L_0 (L_{-1} ^{-1} L_0 )^m \nonumber\\
&=& (mL_{m+n} + L_0 L_{m+n}) - (nL_{m+n}
+L_0 L_{m+n} )\nonumber\\
&=&(m-n)L_{m+n},
\eea
where $m$,$n\geq 0$.  Similarly, one finds that

\bea
[L_m, L_{-1}] &=& L_0 (L_{-1} ^{-1} L_0 )^m L_{-1}
-L_{-1} L_0 (L_{-1} ^{-1} L_0 )^m \nonumber\\
&=& L_0 (L_{-1} ^{-1} L_0 )^{m-1}(1+L_0) -
L_0 ^2 (L_{-1} ^{-1} L_0 )^{m-1} +
L_0 (L_{-1} ^{-1} L_0 )^{m-1} \nonumber\\
&=& 2L_{m-1} +(m-1)L_{m-1}\nonumber\\
&=& (m+1)L_{m-1}.
\eea

It is thus clear that, given a
system of conformal mechanics with $SL(2,R)$ symmetry,
the invertibility of $L_{-1}$ implies the existence of
a half-Virasoro algebra consisting of all Virasoro modes
$L_m$ with $m\geq -1$.  In particular, a scale-invariant
theory will also exhibit conformal invariance (with
$K=-D H^{-1} D$), provided
that there are no zero-energy states.  This can be
seen by defining $L_{-1} =-{\imath \over \hbar} H$ and
$L_0 = -{\imath \over \hbar} D$ and applying the above
results.  Since $H$ is hermitian, $L_{-1}$ will be
invertible if $H$ has no eigenvalues which are zero.
From the above construction one will find a half-Virasoro
algebra with $L_1 = {\imath \over \hbar} K$.

It is not entirely clear
how this statement is reconciled with \cite{MS}, where
it was shown that a scale-invariant Hamiltonian (with
only quadratic momentum dependence) exhibits conformal
invariance only if it admits a closed homothety.
However, \cite{MS} studied
the case where the special conformal symmetry generator
$K$ depended only on the position operators $X$,
and not on
their momentum conjugates.  In the construction given
above, however, it is clear that $K$ will generically
depend on momentum.

In an exactly analogous manner, one can consider the
situation where $L_1$ is invertible (such that one can
consistently define $L_1 ^{-1}$).  In this case, we
make the ansatz
\be
L_{m} = L_0 (L_1 ^{-1} L_0)^{-m} =
(L_0 L_1 ^{-1})^{-m} L_0 \qquad
m\leq 0.
\ee
One finds that these operators yield a half-
Virasoro algebra
\be
[L_m, L_n] = (m-n)L_{m+n}
\ee
which closes for $m\leq 1$.

We might then ask whether it is possible for these two
half-Virasoro algebras to be united into a single full
Virasoro algebra.  The key to this question is the
overlap of these algebras, namely the generators
$L_1$, $L_0$ and $L_{-1}$.  We will demand that these
generators are the same in both half-Virasoro algebras.
This implies the conditions

\be
\label{conditions}
L_0 L_{-1} ^{-1} L_0 = L_1 \qquad
L_0 L_1 ^{-1} L_0 = L_{-1}.
\ee
If $L_0$ is also invertible, then these two conditions are
actually identical.

When (\ref{conditions})
is satisfied, we can actually find
a full Virasoro algebra whose quantum generators are given
by
\bea
\label{defin}
L_m &=& L_0 (L_{-1} ^{-1} L_0)^m =
(L_0 L_{-1} ^{-1})^m L_0 \qquad
m\geq 0 \nonumber\\
L_m &=& L_0 (L_{1} ^{-1} L_0)^{-m} =
(L_0 L_{1} ^{-1})^{-m} L_0 \qquad
m\leq 0
\eea

One would like to show that the generators defined in this
way satisfy the full quantum Virasoro algebra
$[L_m , L_n] = (m-n)L_{m+n}+ f(m)\delta_{mn}$.  For the
case where either $m$,$n \geq 0$ or $m$,$n\leq 0$, the
algebra is obviously satisfied.  Consider the case
$m>0$, $n<0$, $m+n>0$.

\bea
[L_n, L_m] &=& [(L_0 L_1 ^{-1})^{-n} L_0  , L_0
(L_{-1} ^{-1}L_0 )^m ] \nonumber\\
&=& (L_0 L_1 ^{-1} )^{-n-1} L_{-1} L_0 (L_{-1} ^{-1}
L_0 )^m - (L_0 L_{-1} ^{-1})^m L_0 L_{-1}
(L_1 ^{-1} L_0)^{-n-1} \nonumber\\
&=& L_{n+1} L_{m-1} - L_{m-1} L_{n+1} -
(L_0 L_1 ^{-1} )^{-n-1} L_{m-1} \nonumber\\
&\,& - L_{m-1} (L_1 ^{-1} L_0)^{-n-1} \nonumber\\
&=& [L_{n+1}, L_{m-1}] -2L_{m+n}.
\eea
After anchoring the recursion relation with
$[L_{-1}, L_m] = -(m+1)L_{m-1}$,
one finds that the Virasoro algebra
is satisfied for all $m$ and $n$ in the range of
interest.  A similar argument shows that this is
also true for $m+n<0$.  Thus, one sees that any theory
of conformal quantum mechanics which satisfies the
above conditions also has the symmetries of a full
Virasoro algebra.  In addition, it is clear from the
above calculation that the algebra has no central
charge.

Given a theory with operators $L_0$ and
$L_{-1}$ which satisfy the appropriate commutation
relations, where $L_{-1}$ is invertible, one
can simply define $L_{1}=L_0 L_{-1} ^{-1} L_0$.  The
invertiblity of $L_1$ implies the existence of two
half-Virasoro algebras,
and the invertibilty of $L_0$ would further imply
that these two algebras combine to form a single Virasoro
algebra.

Note that, under the (\ref{embedding}), $L_{-1,0,1}$ are
anti-hermitian.  It is clear that if this is the case, then
the $L_m$'s defined by (\ref{defin}) are all
anti-hermitian.  However, in the context of $1+1$ conformal
field theory one usually defines $L_m$'s which satisfy
the hermiticity property $L_n ^{\dagger} = L_{-n}$.  If
this property is satisfied by $L_{-1,0,1}$, then the
$L_m$'s defined by (\ref{defin}) satisfy it as well.

\subsubsection{A simple example}

Consider the Hamiltonian of a non-
relativistic free
particle in one dimension, $H={1\over 2} p^2$ (where
the coordinate has been rescaled in order to absorb the
mass into the conjugate momentum).  If one writes the
standard dilatation operator $D={1\over 4} (rp+pr)$, one
finds that $H$ and $D$ satisfy the standard commutation
relation
\be
[D,H]=\imath \hbar H.
\ee
We may project out of the Hilbert space all states whose
wavefunctions are even under $r\longrightarrow -r$ (this
is, in fact, exactly what we would do if we treated this
as the Hamiltonian for the radial wavefunction of a free
particle in three dimensions with no angular momentum).
The remaining energy
eigenstates have wavefunctions of the form
$\psi (r) = A\sin(kr)$ with energies $E={\hbar^2 k^2 \over
2m}$.  One finds that there is a continuum of eigenstates
with arbitrary positive energy (as scale-invariance demands),
but there is no normalizable
zero-energy state (as the wavefunction
for such a state would vanish everywhere).  Therefore,
one may invert the Hamiltonian and define the operator
\be
K=-D H^{-1} D = -{1\over 2} r^2 - {3\over 8} \hbar^2
{1\over p^2}.
\ee
This operator differs from the usual special conformal
symmetry generator ($K=-{1\over 2} r^2$), but nevertheless
is well-defined and allows the conformal algebra to
close.  By writing $K$ as a momentum-space operator
(substituting $r= \imath \hbar {\partial \over
\partial p}$), one can easily show that $K$ also has no
normalizable eigenstates with zero eigenvalue.  Using the
embedding defined in (\ref{embedding}), one can verify
straightforwardly that the conditions (\ref{conditions})
are satisfied.  This means that one can write the
quantum generators of a full Virasoro algebra
(\ref{defin}).

In many constructions of conformal quantum mechanics
it is common for the operator $H' = {1\over 2} (H-K)$
(in our conventions)
to be used as the Hamiltonian, due to the fact that it
has a discrete spectrum.  In the example given above, one
finds
\be
H' = {1\over 2} (H-K)= {1\over 4}p^2 + {1\over 4}r^2
+ {3\over 16}{\hbar ^2 \over p^2}.
\ee
Under the phase space rotation $\tilde r = p$,
$\tilde p = -r$, we see that this is the
Hamiltonian for a bound particle (indeed, it is a
Hamiltonian of the Calogero-Moser form), and thus
has a discrete spectrum which is bounded from
below.

\section{Relation to $AdS_2$}

\vskip .3in

$\it {\vbox{
    \halign{#\hfil         \cr
           \hskip 1.9in {\it Perhaps there is some vital bit of} \cr
           \hskip 2.1in {\it evidence which eludes us.} \cr
           \hskip 2.2in Belloq, ``Raiders of the Lost Ark" \cr
           }  
      }   
      }$

\vskip .4in

\subsubsection{Calogero models}

The naive $AdS/CFT$ correspondence suggests that there
is a boundary conformal quantum mechanics which is dual
to quantum gravity on the space $AdS_2 \times S^2 \times
T^6$.  In \cite{GibTown}, it was suggested that this
bosonic part of the
boundary conformal quantum mechanics is actually given
by the $N$ particle Calogero model with Hamiltonian
\be
H= {1\over 2} \sum_{i=0} ^N p_i ^2 + \sum_{i<j}
{\lambda \over |r_j - r_i|^2 }.
\ee
It is well-known that this system has conformal symmetry.
It is also well-known that the Hamiltonian for this
system has no ground state.  The results above thus indicate
that the $N$ particle Calogero Model respects
the symmetries of a half-Virasoro algebra.  To determine if
the other half of the Virasoro algebra is also present, one
would have to calculate $L_1 = L_0 L_{-1} ^{-1} L_0$ and
determine its spectrum, a more complicated task which will
not be attempted in this work.

The Calogero model itself is a limit of the more
general Calogero-Moser model, whose Hamiltonian
is given by
\be
H= {1\over 2} \sum_{i=0} ^N p_i ^2 + \sum_{i<j}
{\lambda \over |r_j - r_i|^2 } + {k\over 2}
\sum_{i=0} ^N r_i ^2.
\ee
It is already known \cite{AJ1,AJ2,AJ3,Berg}
that the Calogero-Moser
model exhibits the symmetries of a Virasoro algebra, with
the embedding $L_0 = H$.  Note that this is not the embedding
which we discussed earlier.  In fact, the Calogero-Moser model
is not conformally invariant, as its Hamiltonian does not have
a continuous spectrum.  But in the limit $k \longrightarrow 0$,
the Calogero-Moser model reduces to the Calogero model in
question.  However, in that limit generators of the Virasoro
algebra used in \cite{Berg} become infinite (as they contain
negative powers of the constant $k$).  This is not unexpected
(when $H$ is related to $L_0$) because the Calogero-Moser
model has a discrete spectrum, whereas the Calogero model
has a continuous spectrum.

However, in the construction given by (\ref{embedding})
and (\ref{defin}), the generators
of the half-Virasoro algebra are well-defined.  If the
Calogero model is indeed related to the boundary
conformal theory which is dual to quantum gravity on
$AdS_2 \times S^2 \times T^6$, then it is interesting
to note that it contains at least the symmetries of a
half-Virasoro algebra, whereas it is known that quantum
gravity on $AdS_2$ contains the symmetries of a Virasoro
algebra.

\subsubsection{Probing with a test particle}

One may also consider the quantum mechanics describing
a test-particle in the background of $adS_2$.  The
Hamiltonian for such a particle was discussed in
\cite{CDKKTV}, where it was found that in the non-
relativisitic near-horizon limit the bosonic part of the
Hamiltonian reduced
to that of DFF conformal quantum mechanics \cite{DDF},

\be
H= {p^2 \over 2} + {2J^2 \over r^2}.
\ee
It is clear that this Hamiltonian has no normalizable
zero-energy eigenstates, thus indicating that it
respects the symmetries of a half-Virasoro algebra.
Determining the invertibility of $L_1$ (or equivalently,
$K$) is more complicated, and will not be attempted here.
However, in the case where $J=0$, the system simply
reduces to the free particle example discussed earlier,
where it is clear that $K$ is invertible, and thus that
the other half of the Virasoro algebra can also be
constructed.  One should note that the form of the
quantum generators found here is identical to the form
of the classical generators of the Virasoro algebra found
in \cite{Kumar} for the same system (up to the terms
related to normal ordering).  We see now that
these generators not only generate a classical symmetry
under Poisson brackets, but also generate a full quantum
symmetry under commutators.

\section{Discussion and Further Research}
\vskip .4in

$\it {\vbox{
    \halign{#\hfil         \cr
           \hskip 1.8in Eaton: {\it We have top men working on  }\cr
           \hskip 2.7in {\it it right now. } \cr
           \hskip 1.8in Indy:\hskip .2in {\it Who?} \cr
           \hskip 1.8in Eaton: {\it ... Top men.} \cr
           \hskip 2.1in ``Raiders of the Lost Ark" \cr
           }  
      }   
      }$

\vskip .4in

We have shown that a scale-invariant quantum mechanical system
with no zero-energy states exhibits not only $0+1$ conformal
invariance, but also the symmetries of a half-Virasoro algebra
(defined as the algebra $L_m$ for $m\geq -1$).  If the operator
$L_1$ defined by this half-Virasoro algebra is also
invertible, then the system also exhibits the symmetries of another
half-Virasoro algebra (given by the generators $L_m$ with
$m\leq 1$).  If these two half-Virasoro algebras have identical
generators $L_0$, $L_{-1}$ and $L_1$, then the two half-Virasoro
algebras in fact form a single Virasoro algebra with no central
charge.

It is noteworthy that these constructions of the Virasoro
algebra exhibit zero central charge.
From the $AdS_2/CFT_1$ correspondence, one would expect
superconformal quantum mechanics to be the boundary theory dual
to string theory on $AdS_2 \times S^2 \times T^6$.  The gravity
theory on $AdS_2$ has the symmetries of a Virasoro algebra, as it
is a 2D theory of gravity on a strip.  This Virasoro algebra should
not have a central charge when the effects of ghosts are also
included.  It had been speculated that this might imply that
the dual superconformal quantum
mechanics has the symmetries of a Virasoro algebra with no
central charge.
But in \cite{CKZ}, it was shown that the construction of the
Virasoro algebra given in \cite{Kumar} was actually contained
in a larger $w_{\infty}$ algebra.  Calogero models have also
been shown to exhibit the symmetries of a $w_{\infty}$ algebra;
the properties of these systems were studied extensively in
\cite{AJ1,AJ2,AJ3}.
For the case of a particle in
the background of $AdS_2$, it was shown in \cite{CKZ} that
the central charge associated with the Virasoro algebra of
coordinate diffeomorphisms of $AdS_2$ is replicated by the
unique central extension of the $w_{\infty}$ algebra.
However, our result seems to be more general because
it holds for conformal theories which are not necessarily
connected to $AdS_2$.
But it seems
not unreasonable to expect that a fuller examination of
theories of conformal quantum mechanics of an arbitrary number
of variables will also show that the Virasoro algebras found
here can be extended to $w_{\infty}$ algebras.  If so, perhaps
a study of the central extensions of this $w_{\infty}$ algebra
(and a comparison of this with the central charge of the
diffeomorphism algebra of $AdS_2$) will shed more light on the
connection between $AdS_2$ and conformal quantum mechanics.

There are several directions in which further work can proceed.
It would be very interesting to understand the circumstances under
which (\ref{conditions}) held more systematically.  So far, work
has focused only on the bosonic theory; further investigation of
supersymmetric generalizations of these constructions is required.
It is also important to better understand the physical significance
of the half-Virasoro algebras which have been found here.
Finally, the connections found here between conformal quantum
mechanics and the Virasoro algebra further strengthen the notion
that there is a deep connection between conformal quantum mechanics
and $1+1$ conformal field theory \cite{Andy}.  The relationship between
this connection and the $AdS/CFT$ correspondence should be the
subject of future work.
\vskip .3in
{\bf Acknowledgements}
\vskip .1in

We gratefully acknowledge S. Cullen, S. Kachru,
S. Lee, M. Schulz, S. Shenker,
A. Strominger and N. Konstantine Toumbas
for useful discussions.
We would also like to thank Caltech (where this work was
first conceived) for its hospitality.
This work has been supported by NSF grant
PHY-9870115.

\end{document}